# From the Ground Truth Up:
# Doing AI Ethics from Practice to Principles

## Abstract

Recent AI ethics has focused on applying abstract principles downward to practice. This paper moves in the other direction. Ethical insights are generated from the lived experiences of AI-designers working on tangible human problems, and then cycled upward to influence theoretical debates surrounding these questions: 1) Should AI as trustworthy be sought through explainability, or accurate performance? 2) Should AI be considered trustworthy at all, or is reliability a preferable aim? 3) Should AI ethics be oriented toward establishing protections for users, or toward catalyzing innovation? Specific answers are less significant than the larger demonstration that AI ethics is currently unbalanced toward theoretical principles, and will benefit from increased exposure to grounded practices and dilemmas.

## Keywords

AI ethics, Trustworthy AI, Philosophy and AI, Healthcare AI, AI case studies

## Author

James Brusseau
jbrusseau@pace.edu
1. Philosophy Department, Pace University, New York, USA
2. Signals and Interactive Systems Lab, Department of Information Engineering and Computer Science, University of Trento, Trento, Italy

## 1. Introduction

Artificial intelligence is knowledge produced from pattern recognition, which contrasts with the Kantian vision of human knowledge as created from sequential reasoning. This distinction at – and as – the source of understanding itself means AI cannot be relied upon to obey human conventions of rationality. So, an ethics is needed to domesticate the machines.

Initial work has been largely theoretical, with 84 sets of principles introduced in the last several years, and more on the way. They accumulate along with a challenge. Morley, Floridi, Kinsey and Elhalal (2020) write that these abstract principles urgently require effective translation for application in concrete reality. About that, they are right, but not completely: the gap between theory and practice also implies complementary investigating in the other direction. Instead of descending from principles to practice, these explorations *start* from palpable human experiences and only subsequently climb to abstraction and the theoretical work of comparing and justifying principles.

This report covers the initial phase of that grounded process, the birth of insight and ethical propositions from philosophers, engineers, and medical doctors laboring to make AI work for flesh and blood patients. As for these emergent and speculative propositions, there is no pretension here to defend them against the rules promulgated by today's institutional powers, the ambition is only to demonstrate how ethics can surge from men and women struggling with algorithms and statistics.

## 2. Cases

Going from practice to principles starts with cases. The two discussed here were produced by a team of philosophers, computer scientists, lawyers, and doctors organized out of the Frankfurt Big Data Lab in Germany (Zicari et al. 2021a, Zicari et al. 2021b). Working with algorithmic startup companies, we collaboratively explore their development experiences, with attention split between ethics, technology, law, and medicine (Brusseau 2020).

Our skin lesion case started with a team led by Andreas Dengel and their solution to a debility currently afflicting artificial intelligence diagnoses of skin cancer. Typically, a skin lesion image is analyzed by a neural network to predict whether the lesion is malignant. The procedure is noninvasive, quick, and, in terms of accuracy, machines are now outperforming well-trained dermatologists (Brinker et al. 2019). Still, the technology's use remains limited. The obstacle is the AI blackbox (Lucieri et al. 2020). Doctors may be convinced that the image analysis *generally* works better than their own eyes and experience, but because there is no way to know how the machine reached its conclusion in any particular case, they constantly fear their patient may be an outlier. While aberrations are statistically improbable, there is no escaping the startling errors accompanying advances in image recognition technology. There is even a narrow research area dedicated to provoking comedically wrong outputs, like bananas mistaken for toasters (Brown et al. 2017).

The result is that the barrier to AI-fortified skin health is not technological advance so much as doctors' confidence. And that apprehension converted into a business opening for Dengel and his team. Their Explainable AI in Dermatology product – exAID – wraps around existing artificial intelligence diagnoses and translates the AI method into traditional medical language and reasoning (Lucieri et al. 2020). With the AI processing explained, dermatologists may confirm or reject the mechanical diagnosis in their own human terms, and so conclude with sufficient confidence to recommend action. Technology that was accurate but neglected now becomes practically useful.

The other case starts from cardiac arrest in Denmark, and with a team lead by Stig Nikolaj Blomberg (Blomberg et al. 2021). They responded to an urgent question: Could AI eavesdrop on frantic 112 calls – the European 911 – and perceive humanly imperceptible clues that the subject was suffering cardiac arrest as opposed to some less urgent malady? The information is crucial as cardiac arrest requires specific and immediate treatment, even from bystanders since every minute without resuscitation increases fatality probability by about 10% (Murphy et al. 1994). Dispatchers at Denmark's Emergency Medical Center were failing to identify 25% of the incoming cardiac arrest calls, and so losing precious opportunities to provide the caller with instructions in cardiopulmonary resuscitation (Blomberg et al. 2019).

To save lives, Blomberg's group developed a machine learning tool to detect cardiac arrest and then alert Dispatchers with a light added to their console. When the technology was implemented, it produced unsurprising and also surprising results. Unsurprisingly, true cardiac arrest was recognized more frequently and quickly by the artificial intelligence than by its human partner (Zicari et al. 2021a). However, the machine was also less specific: the AI-returned many false positive alerts which were largely ignored by the dispatchers.  This leads to the surprising result. The dispatchers also largely ignored the true positive alerts. They ignored the AI almost entirely. Dejectedly, Blomberg concluded: "While a machine learning model recognized a significantly greater number of out-of-hospital cardiac arrests than dispatchers alone, this did not translate into improved cardiac arrest recognition by dispatchers" (Blomberg et al. 2021).

### 3. Explainability or Performance?

At the highest level, European AI ethics is dedicated to creating technology that is trustworthy (HLEG 2019). Both cases confronted that trust problem, but the project leaders diverged on the question underneath: *why* believe in algorithmic conclusions in the first place? This uncertainty is not so much about how much confidence exists, but on what the confidence is built. The distinction divided the teams fundamentally.

Dengel's exAID product builds trust from explainability: machine image analysis will be accepted when it is understood. Consequently, his AI wrapper is built to show how skin images are classified, and to verify that the processing functions through disease-related concepts similar to those employed by conventional dermatologists. The key is to translate away from the statistics and probabilities and into seven clinical skin characteristics perceived through close visual inspection. They are: Typical Pigment Network, Atypical Pigment Network, Streaks, Regular Dots and Globules, Irregular Dots and Globules, Blue Whitish Veil (Lucieri et al. 2020, Argenziano et al. 1998). The presence of these telling traits is quantified by exAID, and then overlayed on the image of the skin lesion under scrutiny before being returned to the dermatologist. What is significant here is that the outputted numbers do not *replace* the images with an objective result, instead, they *describe* the images, they direct doctors' attention back to specific visual evidence so they can check for themselves. Without the exAID wrapper, all a dermatologist receives is a cold numerical probability sent back in return for submitted images. With the wrapper, the statistics no longer substitute skin pictures, they enhance them.

In essence, machine learning reverses: instead of the material world converting into a digital score, the score guides a way back into material experience. Numbers serve eyes, not the other way around.

The original skin lesion picture also receives a second overlay, a layer of color indicating exactly where the AI detected the information resulting in diagnosis. So, the doctor learns not only of a blue whitish veil, but precisely where on the lesion it can be found. Here again, explaining the AI does not mean adding still more digital information or another set of statistical methods (Shapley values or similar) to approximate which pieces of data contributed how much to prior statistical processing (Chen 2021). Instead, explaining means translating the experience of interacting with the system into familiar and human modes of seeing, locating, touching. The machine is humanized, anthropomorphized.

In Denmark, a different strategy: instead of understanding, Blomberg and his team leveraged power. AI performance *imposed* trust. To force dispatcher respect for the cardiac arrest alerts, the machine was tuned to defeat humans. While it is true that an abundance of false positive alerts got logged along the way, the hard fact remained that true cardiac arrest was detected faster by algorithms and data than by human listening and experience.

The idea of helping the human dispatchers catch up with the AI instead of leaving them behind – perhaps by developing software to clarify or augment the audio the dispatchers heard – was never broached in discussions with our group. Just the opposite, we learned that as the AI progressed, humanity reduced. Originally, the AI processed the raw audio of calls, thick with their anguish. However, the discovery was made that background screams and lamentations were major sources of false positive alerts, and so a two-stage approach was developed. An initial filter eliminated human emotion by transcribing the calls into dry words, and then a second process analyzed the text for patterns in vocabulary, in sentences, in questions and answers, and in specific, described characteristics. Blomberg explained that if the caller states that the subject is unconscious, then the probability of cardiac arrest rises. If blue lips are mentioned, the probability also rises. If both, the alarm illuminates (Zicari et al. 2021a). However, beyond that and a few similar anecdotes, there was no human-oriented discussion of the AI process, nothing that would make sense to a doctor. There were only the statistical outcomes of sensitivity (the detection of cardiac arrest) and specificity (the ratio of true cardiac arrests detected, against false alerts), and how they could be improved.

Though the model was partially open-sourced (Havtorn et al. 2020; Maaløe et al. 2019), our group did not pursue a technical understanding of explaining how the system worked because we were convinced that winning human trust for the AI decisions would originate and remain within the parameters of performance defined as speed and accuracy. Ultimately, the objectivity – the mathematical certainties – were not just descriptions of functionality but conceptual rigidities that commanded respect by humiliating human subjectivity and uncertainty. While emergency call dispatchers considered and doubted and floundered and let seconds pass through their indecision, hard numbers responded. The result for the emergency call AI technology was a kind of trust not won from dispatchers so much as stamped onto them. As opposed to the previous case which drew human explanations from an inanimate machine, here, the power of inanimate machines was programmed to crush human doubts.

In actual practice in Copenhagen the dispatchers were not crushed, they resisted by ignoring the technological prompts. But that failure does not change the nature of the strategy, it only requires still more engineering and perfecting.

Which of the two paths to trustworthy AI is recommendable? Is it explainability so that the machine will be *understood* well, or accurate performance so the machine will *work* well? The case for explainability flourished in German dermatologists' offices: it was because the skin analyzing machine was finally understood well, that it was allowed to work well.

In the Danish emergency call center, a different choice was unavoidable. Faced with the reality that dispatchers were ignoring cardiac arrest alerts, Blomberg could only tune his machine still tighter, and press his mechanical advantage still harder because he knew that death hung on seconds. This is the stark reality he faced: No time to draw attention to keywords, or to describe what was revealing in sentence patterns. No time to explain what linguistic cues triggered an alert, or even to determine whether there could be any explanation. The AI's flashing light could only present dispatchers a decision to be made instantaneously, or to be made for them while they hesitated. In that way, cardiac arrest telephone calls resemble one-way airplane tickets and romantic passions and so many of the reasons we want to be alive in the first place: if you stop to ask why, it is already too late.

For his part, Dengel's skin lesion team reminded our group of the General Data Protection Regulation stipulation that data subjects possess a *right to explanation* for any automated decision made by computer algorithms (Lucieri 2020). That will have to change, though, because no one has the responsibility to try the impossible, and the cardiac arrest case demonstrates that under time's pressure and on the edge splitting life from death the only functional source for trust in AI technology is power as demonstrated by performance. The machine is faster than humans. Which means that the proposal of a *right* to an explanation for an algorithmically generated decision is cancelled by the demands of life itself – not just that patients stay alive, but that life with AI is worth living.

Still, the question remains: Where does the line get drawn? In which instances should trust be built on humanized knowledge about AI decisions? And, when should trust be sought through algorithmic power? More cases will need to be studied, but what these two indicate is that when the moment is critical and the risk is high, power is better than knowledge.

### 4. Trustworthy or Reliable?

Trustworthy AI is the titular goal of European Commission publications on AI ethics, but should it be? Joanna Bryson (2018) and Mark Ryan (2020) advocate for a shift toward inanimate reliability, and that transition gains support from ground-up work, from investigating that begins from sincere psychological attitudes and bare human experiences.

This is bare experience: trust is inseparable from betrayal. If there was no dishonesty or infidelity than we would not need the concept of trustworthiness. We would not even have it, we could not have it. As Derrida (1998) has demonstrated, these kinds of dialectic word pairings are not just opposites or contrasts, they are episodes of co-dependence. Each requires the other in order to produce linguistic meaning. The simplest example may be honesty and lies: if no one ever told the truth, it would be impossible to invent the word or even the idea of lying. This paradox explains why no one calls actors liars even while we are absorbed in the lies their characters tell. And it explains the peculiar linguistic experience called bullshitting according to Frankfurt (2005): words and claims that are neither truths nor lies, just absurdities.

In lived experience, the fundamental distinction is not between trust on one side and betrayal on the other. Instead, it is between a reality of trust entwined with betrayal on one side, and other realities without either one. As machines are enveloped in language, they too are subjected to the distinction: both or neither.

It is easy to write that trustworthiness is attractive and that deceit is repellent, but judging from how we live, any neutral observer would conclude the opposite: deception and dishonesty are alluring. The evidence is everywhere, but most immediately in our movies and literature, in the places where we *choose* to spend our time. We are drawn to infidelity while scrolling Netflix, we seek deceit while standing in front of the bookracks at the airport. And if these quotidian examples are too crass, then there is the arousal Shakespeare summons from his audience as Brutus plunges his dagger into Caesar's back. It is not a psychotic delight in blood, but the thrill of betrayal captured in Caesar's recognition that it was Brutus too, not just callous assassins driven impersonally by thirst for power. All of this, finally, is *inseparable* from trustworthiness, it is the way the word and concept gain meaning. And all those who venerate the trustworthy are equally engaged by the dark complements, whether they admit it or not.

It follows that if machines are going to be trusted, if we are going to talk and write about them that way, there is a requirement – a condition of the possibility of *being* trustworthy – that they also betray. If the machine wins our trust when it works well, then the machine's failures are not sites of error so much as scenes of unfaithfulness. The AI that falsely signals a cardiac arrest to an emergency dispatcher is not wrong, it is *duplicitous*, and the result should not be disappointment, but *guilt*.

That never happens, though. Not remotely. In our group's extensive work with two very different startups trying to promote trustworthy AI, not once did a single ethicist, engineer, lawyer, or doctor define false outputs as dishonesty, infidelity, deception, lies, betrayal. No one invoked the idea of being *tempted* by a false positive. It occurred to no one to propose that the light on the emergency call dashboard was winking, trying to lure the dispatcher's attention and seduce with insincere promises. When the machine was wrong, it was just wrong, that was all.

Ultimately, the problem with trustworthy AI – and the reason the idea should be abandoned for the neutral and inanimate term of reliability – is not that we cannot force ourselves as humans to trust the machines. Probably, we can. The problem is what waits on the other side of that trust: mechanical duplicity. So, even if we grant that algorithms *could* be trustworthy, a complete understanding of what that means requires that we also acknowledge engaging with the acidic joys of betrayal. For the human experience of encountering AI today, that joy is inconceivable.

Finally, and stated positively, the way we speak in the real world when machines fail *dictates* the way we must respond when they succeed. With failure, we feel disappointment, and we talk about error and incorrect outputs. We speak about the opposite of reliability. With success, consequently the decision is already made: AI can be reliable, but not trustworthy.

## 5. Protection or Performance?

A sentence with jarring implications appears near the end of our group's report on the Copenhagen emergency call case:

> Under the forthcoming Medical Device Regulation in the EU, the AI system will be classified as medical device, and it would therefore need EC-certification (Zicari et al. 2021a).

Blomberg would not have been able to initiate his experiment today. Medically certifying the AI would swamp development in regulations, permissions, approvals, and ethical safeguards. One obstacle would be the General Data Protection Regulation, and for good reason: the telephone calls are tortured. A victim collapses in agonal wheezing, loved ones frantically seek a reason as the lips tint blue. No one knows what to do and gawking onlookers deepen the helplessness. In the disorienting environment of cardiac arrest, some people will lose control over themselves. Others will maintain control but have no idea how to exhibit their own identity. Either way, regulatory protections promulgated for personal information in the ethical region of human dignity were crafted for these moments. This is when privacy matters. If tragic emergency calls are not shielded from the commodification of machine learning, it is difficult to imagine what human experience could possibly be considered protected.

But commodification works. The machine recognizes cardiac arrest and lives can be saved.

The dilemma is bottomless, and another example of why seeking absolute answers to legitimate ethical questions is misguided. Genuine ethics aims for the limit where it is simultaneously true that it is impossible to decide, and a decision must be made. Blomberg was there, faced the impossibility, and decided. AI development proceeded.

In 2019, Tesla disabled major components of its Autopilot feature in nations where UN/ECE r79 vehicle safety regulations were promulgated (Lambert 2019). The restrictions embodied a commitment to safeguard against artificial intelligence harms and risks, even at the cost of advances in AI development and application (Roberts 2021: 1). They also foreshadowed the regulatory paradox that Blomberg initially escaped, but now cannot. More regulation equals less regulating because innovation is stymied until there is nothing left to restrict.

It does not need to be that way. AI ethics can be reconceived to catalyze innovation. The direct strategy is to credit ethically – not just technically and economically – AI that performs well. The proposal is that performance *is* a value, and it emerges for this reason. One of the most frustrating aspects of our group's interactions with Dengel's skin lesion team and Blomberg's cardiac arrest group was that we found no way to unambiguously account for their pure engineering accomplishments. Instrumental value was easy to locate: both technologies are worth having for the indirect reason that they save lives. There is something more than that, though. The skin lesion explainability wrapper and the cardiac arrest language filter each stand on their own – without regard for their human benefits – as small but identifiable triumphs of design. Some AI engineers grasp this with the concept of code quality (Spinellis 2008), but regardless of the term, in our group's experience it explained the disjunction between AI doing, and what is done. Performance runs along that first vector, not the second, and along the performance scale elevating levels of quality become increasingly rewarding regardless of what is done. In AI ethics from the ground up, pure performance is a value because algorithmic designers find that high quality is worth having on its own.

There is a parallel here with functionalist architecture, like the Bauhaus or Pompidou Center. For those buildings, functionality transcends its native role and becomes form. The buildings' inner workings – the stairways and waterpipes – *are* its decoration. Analogously for AI that functions well, the practical transcends its instrumental role and becomes an end in itself. So, AI that is worth doing because of *what* it does well (serves doctors and patients) becomes AI that is worth doing because it *does* well. The ethical value of performance is intrinsic.

Weighing performance in a vacuum is dangerous. There is no end to the human catastrophes occasioned by exquisitely functioning machines, and advocating pure performance as an ethical value requires acknowledging this: even death-mechanisms like nuclear bombs can be majestic accomplishments, awesome, with all that word implies. The best descriptor may be Kant's sense of the sublime, the feeling of reason's power and superiority over nature (Kant 1987: §28). In this way, performance as a value resembles philosophical approaches to painted art: there is a transcendent dimension to its existence, one that escapes the chemical composition of the oils applied to the canvas, and the money generated or lost, and the history of the work's transfer from one to another owner. There is an escape from the effects the work produces as a material object.

It is difficult to credit that escape with today's conventional AI ethics. The escape is particularly elusive for those working inside the European Commission's *Ethics Guidelines for Trustworthy AI*. The *Guidelines* does include the value of accuracy, but it is buried deep in the document, underneath the four pillars and then within the category of Technical Robustness and Safety (HLEG 2019: 17). Defined dryly as the "ability to make correct judgements," the idea of simple accuracy is as narrow as it is desiccated. Performance is expansive. In the cardiac arrest case, quickness was as critical as accuracy. In the skin lesion case, the output's elegance was as encouraging as the raw calculations of correct predictions. AI that performs well is not only correct and accurate but agile and graceful.

As drawn from our working experience, the four traits of performance as a value in AI ethics are:
- Intrinsically valuable: The only justification required is engineering excellence.
- Independent: The technology is evaluated without regard for its material effects on humans, or the world.

- Expansive: The ingredients of performance – what counts as success – cannot be known, defined, or checklisted beforehand any more than the qualities of venerated art can be named before the work is done. The dimensions of accomplishment are only appreciated after the culmination.
- Balancing: Performance can counter ethical shortcomings. An AI tool that performs flawlessly, for example, does not need to be explainable or transparent.

The central significance of performance as a value is that it transforms the role of AI ethics: it fundamentally reorients mainstream approaches as they are represented by the EC *Ethics Guidelines*. The *Guidelines* places the burden on developers to show that their products are trustworthy, which implies fulfilling requirements including those established to protect personally identifying information. Only then will innovations like Blomberg's emergency call AI be approved for work out in the world. The addition of Performance – accuracy, speed, elegance – to the first line of AI ethics can reverse the priorities. If a machine functions sufficiently well, then it is the regulators who carry the burden of showing why the technology should be constrained. Because creative engineering is understood as *intrinsically* good and worth pursuing, the burden for justifying restrictions falls toward the restrictors.

Instead of the creators proving themselves to regulators, now it is regulators who must prove to creators.

In some cases, the proving will be easy and the addition of performance to AI ethics imperceptible. Take the example of social media recommendation algorithms attuned to maximize users' angers and frustration. In that case, like so many others, the ethics of individual human autonomy and collective social welfare easily overcomes the value of the predictive analytics accomplishment. Just because the machine works does not mean it should be built. In close calls, however, in cases resembling the cardiac arrest AI where real human dangers weigh against significant benefits, the accomplishment of the machine itself may prove decisive. If there is no way to be certain beforehand whether an AI ultimately helps or harms humanity, and if the technology performs, then that value endorses and potentially justifies pushing ahead, through the unknown.

Ultimately, there are two conceptions of AI ethics, one without and the other with performance as a value. One postures defensively against risks. The other leans forward to catalyze opportunities. There will be no way to know which is preferable until it is too late, just as there are no obviously right or wrong places to draw the line between AI ethics as protecting humanity from innovation, and AI ethics as stimulating more of it. But there are different places.

## 6. Conclusion

In the beginning of philosophy Plato taught that ethics was about purification: knowledge became truer as it increased in abstraction and decreased in humanity. The light-headedness of intellectual exploration was better than the buzz of wine, the desire for mathematics more passionate than sex. The premise of this paper is that Platonic urges have limited AI ethics. Overbalancing principles against practice forces us to believe that we only need to get the abstract, sterile theory right, and then application in the slippery world will come as an afterthought.

This essay is an exercise in reversing the thinking: instead of learning about the world by escaping upward from it, knowledge is produced by diving down into the work of real AI developers as they crash into human dilemmas. Readers will draw varied conclusions about explainability and performance, and about trustworthiness and reliability, and about the intrinsic value of innovation, but the underlying assertion is that those conclusions will emerge stronger from the ground of lived human experiences than from cloistered philosophy.

## Declaration

Author declares that there is no conflict of interest or external funding.

End